\documentclass[]{article}

\usepackage[margin=0.75in]{geometry}

\usepackage[utf8]{inputenc}

\usepackage{hhline}
\usepackage{multirow}

\usepackage{subcaption}

\usepackage{amsmath, amsthm}
\usepackage{amssymb,amsfonts}

\usepackage{graphicx}
\usepackage{comment}

\usepackage{color}

\usepackage[hidelinks]{hyperref}

\usepackage{array} 

\newcommand{\sm}{\textrm{-}}
\newcommand{\smi}{{\sm i}}

\renewcommand{\epsilon}{\varepsilon}

\newcommand{\proj}{\mathrm{proj}}
\newcommand{\sens}{\mathrm{sens}}

\newtheorem{lemma}{Lemma}

\newtheorem{definition}{Definition}

\newcommand{\bi}{\begin{itemize}}
\newcommand{\ei}{\end{itemize}}
\newcommand{\ben}{\begin{enumerate}}
\newcommand{\een}{\end{enumerate}}
\newcommand{\be}{\begin{equation}}
\newcommand{\ee}{\end{equation}}
\newcommand{\ba}{\begin{array}}
\newcommand{\ea}{\end{array}}

\title{Shapley-Based Core-Selecting Payment Rules}
\date{First version: July 2, 2021\\This version: July 2, 2021}

\author{
Vitor Bosshard, Department of Informatics, University of Zurich\\
Sven Seuken, Department of Informatics, University of Zurich and ETH AI Center\\
\{bosshard,seuken\}@ifi.uzh.ch
}

\begin{document}

\maketitle


\begin{abstract}
In this research note, we lay some groundwork for analyzing the manipulability of core-selecting payment rules in combinatorial auctions.
In particular, we focus on payment rules based on the bidders' Shapley values.
We define a sensitivity metric, and provide analytical results for this metric in LLG, for six different payment vectors used as reference points for minimum-revenue core-selecting payment rules.
We furthermore show how this sensitivity affects the derivative of the resulting payment rules.

\end{abstract}



\section{Introduction}

Combinatorial auctions (CAs) are used in settings where heterogeneous, indivisible goods are auctioned off to multiple bidders, allowing them to express non-linearities in their valuations of bundles of goods.
The strategyproof VCG mechanism has many undesirable properties in this setting \cite{ausubel2006lovely}, so finding a CA payment rule that leads to good auction outcomes is an open problem.
Rules that are not strategyproof are most often used in practice, but predicting the equilibria they produce is not straightforward \cite{bosshard2020JAIR}.
If we could quantify how manipulable different rules are, this would enable us to make proper tradeoffs between manipulability, revenue, and other desiderata, and also propose new rules with better properties along those dimensions.
We suspect that the partial derivatives of a payment rule are good indicators of manipulability and can thus be used to estimate the auction's equilibrium outcomes, a link we aim to establish more precisely in future work.

In this research note, we focus on \emph{reference point rules} \cite{lubin2018designing}, a class of payment rules that is built by choosing a reference point in the space of possible payments, and then projecting onto the \emph{minimum revenue core} from there to derive the final payments.
We define a \emph{sensitivity} of each reference point, and show that this metric can be used to determine the derivative of the corresponding reference point rule in the LLG domain.
We investigate several such rules, finding that a reference point based on bidders' Shapley values in the coalitional value game \cite{shapley1953value} is less sensitive than the widely used VCG-nearest rule.
The implied low manipulability of this rule is consistent with previous findings based on numerical simulation \cite{lubin2018designing}.

\section{Preliminaries}
\label{sec:fm}

\subsection{Formal Model}

A combinatorial auction (CA) is a mechanism used to sell a set $M = \{1, 2, \ldots, m \}$ of goods to a set $N = \{1, 2, \ldots, n \}$ of bidders.
Each bidder $i$'s preferences over bundles are captured via the bidder's \emph{valuation} $v_i$. Specifically, for each bundle of goods $K \subseteq M$, we let $v_i(K) \in \mathbb{R}_{\geq 0}$ denote bidder $i$'s  value for bundle $K$. 
%

Each bidder submits a bid $b_i$ to the auction, which is a (possibly non-truthful) declaration of her whole valuation.
The bid profile $b = (b_1, \ldots, b_n)$ is the vector of all bids from all bidders.
The bid profile of every bidder except bidder $i$ is denoted $b_{\smi}$.
Similarly with $b_{-K}$ for a set of bidders $K$.
We also express the combination of multiple bid profiles $b$ and $b'$, by $(b_K, b'_{-K})$, where each bidder's bid is taken from $b$ resp. $b'$.

For an arbitrary subset $K$ of bidders, we define the coalitional value $V_K$ as
$$V_K(b) := \sum_{j \in K} b_j(X_j(b_K, 0_{-K})),$$
i.e., the joint value this set of bidders would obtain from the goods if all other bidders were excluded from the auction (or equivalently, their bids set to $0$).
Thus, $V_N$ is the social welfare obtainable under $b$.

The CA has an allocation rule $X$, determining an allocation $x=X(b)$, where $x_i = X_i(b)$ denotes the bundle assigned to bidder $i$. 
We require the resulting allocation to be \emph{feasible},  i.e., $\forall i,j \in N: x_i \cap x_j = \emptyset$, and we let $\mathcal{A}$ denote the set of all feasible allocations. We only consider  \emph{efficient} allocation rules, i.e., rules that produce an allocation that maximizes \emph{reported social welfare} (which is the sum of bidders' reported values).
The CA also has a payment rule $p$ which is a function assigning a payment $p_i(b) \in \mathbb{R}_{\geq 0}$ to each bidder $i$.
Together, the allocation $x$ and the payment vector $p$ are called the auction \emph{outcome}.

\subsection{The Core}

Informally, a payment vector $p$ is said to be \emph{outside the} \emph{core} if a coalition of bidders is willing to pay more for the goods than what the mechanism currently charges the winners. To avoid such outcomes, Day and Milgrom (2008) introduced the idea of \emph{core-selecting payment} rules that restrict payments to be in the core.

To define the core, we will need to define the welfare achieved by a subset of bidders $K$, when goods are allocated efficiently among all bidders (and not just among the subset $K$ as was the case with $V_K$):
$$W_K(b) := \sum_{j \in K} b_j(X_j(b)).$$

\begin{definition}
Given allocation rule $X$ and a bid profile $b$, an auction outcome is in the \emph{core} if $X$ is efficient and the payments $p(b)$ satisfy the individual rationality constraints and the following additional core constraints:
\begin{equation}
    \forall L \subseteq N:
    \sum_{j \in N \setminus L} p_j(b) \geq V_L(b) - W_L(b).
    \label{eq:core}
\end{equation}
\end{definition}
Each of the sets $L$ is called a \emph{blocking coalition}, and it requires that the joint payment of the remaining bidders $N \setminus L$ is at least as large as the joint loss of welfare incurred by $L$ due to the presence of $N \setminus L$.
Among all payment vectors in the core, the \emph{minimum revenue core (MRC)} is the set of payment vectors with smallest $L_1$ norm, i.e., which minimize the sum of the  payments of all bidders.

\subsection{The LLG Domain}

In this note, we restrict our attention to the Local-Local-Global (LLG) domain \cite{ausubel2006lovely}.
LLG is one of the smallest examples of an auction where combinatorial interactions between bidders arise.
There are three bidders, with bidders 1 and 2 being local, interested in two different single goods, and bidder 3 being global, interested in the package of both goods.

In this note, we do not consider the possibility of the local bidders engaging in \emph{complex bidding}, i.e., making an additional bundle bid on the global bundle $\{1,2\}$, even though such manipulations have shown to be beneficial \cite{beck2013incentives,bosshard2020cost}.
Instead, we assume that each bidder only bids on their bundle of interest.
Thus, we will keep the notation simple and denote the bid profile $b$ as a vector $(A, B, G)$, where $A := b_1(\{1\}), B := b_2(\{2\})$, and $G := b_3(\{1,2\})$.

The global bidder is known to have a dominant strategy to bid truthfully under most payment rules, namely those who are minimum revenue core selecting \cite{beck2013incentives}.
Thus, we will focus our analysis on the two local bidders.

\subsection{Definitions of Reference Points}

We next introduce a series of payment rules.
While some of these rules are used as-is in practice, in this note we consider them as reference points from which a projection to the minimum-revenue core is done.
This method of contructing payment rules has been proposed as a promising design approach in the literature \cite{lubin2018designing}.

\begin{definition}[First Price]
Given a bid vector $b$, the first price payment rule is given by
$$p_i(b) := b_i(X_i(b)).$$
\end{definition}
\begin{definition}[VCG]
Given a bid vector $b$, the VCG payment rule is given by
$$p_i(b) := V_{N \setminus \{i\}}(b) - W_{N \setminus \{i\}}(b).$$
\end{definition}

\begin{definition}[Shapley payoffs/payments without Auctioneer]
Given a bid vector $b$, the Shapley payoffs (without auctioneer) are given by
$$\pi_i(b) := \sum_{S \subseteq N \setminus \{i\}} \frac{|S|! \cdot (|N| - |S| - 1)!}{|N|!} (V_{S \cup \{i\}}(b) - V_S(b)),$$
and the Shapley payments (without auctioneer) are given by
$$p_i(b) := v_i(X_i(b)) - \pi_i(b).$$
\end{definition}

\begin{definition}[Shapley payoffs/payments with Auctioneer]
Given a bid vector $b$, the Shapley payoffs (with auctioneer) are given by
$$\pi_i(b) := \sum_{S \subseteq N \setminus \{i\}} \frac{(|S|+1)! \cdot (|N| - |S| - 1)!}{(|N|+1)!} (V_{S \cup \{i\}}(b) - V_S(b)),$$
and the Shapley payments (with auctioneer) are given by
$$p_i(b) := v_i(X_i(b)) - \pi_i(b).$$
\end{definition}

The only difference between the last two rules are the weights assigned to each coalition. 
This is because the latter considers the auctioneer as an agent in the coalitional value game on which the definition of Shapley payoffs is based.
Clearly, when the auctioneer is not part of the coalition, the value achieved is 0.
Thus, in the latter version of the Shapley payoffs, the set $S$ should be thought of as always including the auctioneer.
The half of subsets for which this doesn't hold can be excluded from the sum, as they don't contribute anything to bidders' payoffs.
We believe that the Shapley payments without seller is the more natural variant, as it has the desirable property that it always lies below the MRC, and never above it, at least in LLG.


\section{Results}

\subsection{Sensitivity of Reference Points}

We used Python with the symbolic solver Sympy \cite{meurer2017sympy} to compute expressions for the various reference points of interest defined above.
Our code is freely available and can be found at\\ 
(\href{https://github.com/marketdesignresearch/Shapley\_MRC}{\color{blue} https://github.com/marketdesignresearch/Shapley\_MRC}).
To perform our analysis in a clean way, we split the set of possible bid profiles $b = (A,B,G)$ into four cases:
\ben
\item Both local bidders make weak bids: $A < G$, $B < G$,
\item The first local bidder makes a strong bid: $B < G < A$,
\item The second local bidder makes a strong bid: $A < G < B$,
\item Both local bidders make strong bids: $G < A$, $G < B$.
\een
Between these four cases lie the boundaries where the optimal allocations of some subsets of bidders change.
The payments change smoothly over those boundaries, but the coalitional value functions that make up the VCG and Shapley payment rules take on different functional forms on different sides of the boundary.

The analytical expressions for the reference points are given in Table~\ref{tab:payments}.

In this note, we study a quantity that we denote the \emph{sensitivity} of a reference point.

\begin{definition}
    Let $p = p(b)$ be a payment rule.
    In the LLG domain, the \emph{sensitivity} of $p$ to bidder 1's bid $A$ is given by
    $$\sens_1(p(b)) := \frac{\partial p_1(b)}{\partial A} - \frac{\partial p_2(b)}{\partial A}.$$
\end{definition}

We define this in terms of bidder 1, with the definition for bidder 2 obviously being symmetric.
The sensitivities of the reference points we study are given as analytical expressions in Table~\ref{tab:payments} as well.
These quantities determine how core-selecting rules using this reference point behave, as will become clear in the next subsection.
%
\renewcommand{\arraystretch}{1.1}
\setlength\extrarowheight{5pt}

\begin{table}
\setlength\tabcolsep{2pt}
\resizebox{\textwidth}{!}{%

\begin{tabular}{|l|l|l|l|l|l|l|l|}
\hline
\multicolumn{2}{|l|}{\bf Case} & \bf \multirow{2}{1cm}{First\\Price}& \bf \multirow{2}{0.6cm}{VCG}& \bf \multirow{2}{3.1cm}{Shapley payment\\w/o seller}& \bf \multirow{2}{2.8cm}{Shapley payoff\\w/o seller}& \bf \multirow{2}{3.1cm}{Shapley payment\\w/ seller}& \bf \multirow{2}{2.8cm}{Shapley payoff\\w/ seller}\\
\multicolumn{2}{|l|}{} &&&&&&\\
\hhline{:=:=:=:=:=:=:=:=:}
\multirow{4}{1.2cm}{\small locals\\weak} & \small $p_1$ & \small $ \displaystyle A$& \small $ \displaystyle - B + G$& \small $ \displaystyle \frac{A}{6} - \frac{B}{3} + \frac{G}{3}$& \small $ \displaystyle \frac{5 \cdot A}{6} + \frac{B}{3} - \frac{G}{3}$& \small $ \displaystyle \frac{7 \cdot A}{12} - \frac{B}{4} + \frac{G}{4}$& \small $ \displaystyle \frac{5 \cdot A}{12} + \frac{B}{4} - \frac{G}{4}$\\[5pt]
& \small $p_2$ & \small $ \displaystyle B$& \small $ \displaystyle - A + G$& \small $ \displaystyle - \frac{A}{3} + \frac{B}{6} + \frac{G}{3}$& \small $ \displaystyle \frac{A}{3} + \frac{5 \cdot B}{6} - \frac{G}{3}$& \small $ \displaystyle - \frac{A}{4} + \frac{7 \cdot B}{12} + \frac{G}{4}$& \small $ \displaystyle \frac{A}{4} + \frac{5 \cdot B}{12} - \frac{G}{4}$\\[5pt]
& \small $\frac{\partial}{\partial A}$ & \small $ \displaystyle \left( 1, 0\right) $& \small $ \displaystyle \left( 0, -1\right) $& \small $ \displaystyle \left( \frac{1}{6}, - \frac{1}{3}\right) $& \small $ \displaystyle \left( \frac{5}{6}, \frac{1}{3}\right) $& \small $ \displaystyle \left( \frac{7}{12}, - \frac{1}{4}\right) $& \small $ \displaystyle \left( \frac{5}{12}, \frac{1}{4}\right) $\\[5pt]
& \small $sens_1(p(b))$ & \small $ \displaystyle 1$& \small $ \displaystyle 1$& \small $ \displaystyle \frac{1}{2}$& \small $ \displaystyle \frac{1}{2}$& \small $ \displaystyle \frac{5}{6}$& \small $ \displaystyle \frac{1}{6}$\\[5pt]
\hline
\multirow{4}{1.2cm}{\small local 1\\strong} & \small $p_1$ & \small $ \displaystyle A$& \small $ \displaystyle - B + G$& \small $ \displaystyle - \frac{B}{3} + \frac{G}{2}$& \small $ \displaystyle A + \frac{B}{3} - \frac{G}{2}$& \small $ \displaystyle \frac{A}{2} - \frac{B}{4} + \frac{G}{3}$& \small $ \displaystyle \frac{A}{2} + \frac{B}{4} - \frac{G}{3}$\\[5pt]
& \small $p_2$ & \small $ \displaystyle B$& \small $ \displaystyle 0$& \small $ \displaystyle \frac{B}{6}$& \small $ \displaystyle \frac{5 \cdot B}{6}$& \small $ \displaystyle \frac{7 \cdot B}{12}$& \small $ \displaystyle \frac{5 \cdot B}{12}$\\[5pt]
& \small $\frac{\partial}{\partial A}$ & \small $ \displaystyle \left( 1, 0\right) $& \small $ \displaystyle \left( 0, 0\right) $& \small $ \displaystyle \left( 0, 0\right) $& \small $ \displaystyle \left( 1, 0\right) $& \small $ \displaystyle \left( \frac{1}{2}, 0\right) $& \small $ \displaystyle \left( \frac{1}{2}, 0\right) $\\[5pt]
& \small $sens_1(p(b))$ & \small $ \displaystyle 1$& \small $ \displaystyle 0$& \small $ \displaystyle 0$& \small $ \displaystyle 1$& \small $ \displaystyle \frac{1}{2}$& \small $ \displaystyle \frac{1}{2}$\\[5pt]
\hline
\multirow{4}{1.2cm}{\small local 2\\strong} & \small $p_1$ & \small $ \displaystyle A$& \small $ \displaystyle 0$& \small $ \displaystyle \frac{A}{6}$& \small $ \displaystyle \frac{5 \cdot A}{6}$& \small $ \displaystyle \frac{7 \cdot A}{12}$& \small $ \displaystyle \frac{5 \cdot A}{12}$\\[5pt]
& \small $p_2$ & \small $ \displaystyle B$& \small $ \displaystyle - A + G$& \small $ \displaystyle - \frac{A}{3} + \frac{G}{2}$& \small $ \displaystyle \frac{A}{3} + B - \frac{G}{2}$& \small $ \displaystyle - \frac{A}{4} + \frac{B}{2} + \frac{G}{3}$& \small $ \displaystyle \frac{A}{4} + \frac{B}{2} - \frac{G}{3}$\\[5pt]
& \small $\frac{\partial}{\partial A}$ & \small $ \displaystyle \left( 1, 0\right) $& \small $ \displaystyle \left( 0, -1\right) $& \small $ \displaystyle \left( \frac{1}{6}, - \frac{1}{3}\right) $& \small $ \displaystyle \left( \frac{5}{6}, \frac{1}{3}\right) $& \small $ \displaystyle \left( \frac{7}{12}, - \frac{1}{4}\right) $& \small $ \displaystyle \left( \frac{5}{12}, \frac{1}{4}\right) $\\[5pt]
& \small $sens_1(p(b))$ & \small $ \displaystyle 1$& \small $ \displaystyle 1$& \small $ \displaystyle \frac{1}{2}$& \small $ \displaystyle \frac{1}{2}$& \small $ \displaystyle \frac{5}{6}$& \small $ \displaystyle \frac{1}{6}$\\[5pt]
\hline
\multirow{4}{1.2cm}{\small locals\\strong} & \small $p_1$ & \small $ \displaystyle A$& \small $ \displaystyle 0$& \small $ \displaystyle \frac{G}{6}$& \small $ \displaystyle A - \frac{G}{6}$& \small $ \displaystyle \frac{A}{2} + \frac{G}{12}$& \small $ \displaystyle \frac{A}{2} - \frac{G}{12}$\\[5pt]
& \small $p_2$ & \small $ \displaystyle B$& \small $ \displaystyle 0$& \small $ \displaystyle \frac{G}{6}$& \small $ \displaystyle B - \frac{G}{6}$& \small $ \displaystyle \frac{B}{2} + \frac{G}{12}$& \small $ \displaystyle \frac{B}{2} - \frac{G}{12}$\\[5pt]
& \small $\frac{\partial}{\partial A}$ & \small $ \displaystyle \left( 1, 0\right) $& \small $ \displaystyle \left( 0, 0\right) $& \small $ \displaystyle \left( 0, 0\right) $& \small $ \displaystyle \left( 1, 0\right) $& \small $ \displaystyle \left( \frac{1}{2}, 0\right) $& \small $ \displaystyle \left( \frac{1}{2}, 0\right) $\\[5pt]
& \small $sens_1(p(b))$ & \small $ \displaystyle 1$& \small $ \displaystyle 0$& \small $ \displaystyle 0$& \small $ \displaystyle 1$& \small $ \displaystyle \frac{1}{2}$& \small $ \displaystyle \frac{1}{2}$\\[5pt]
\hline
\end{tabular}

}
\caption{Analytical expressions in LLG for six different payment rules used as reference points. The analysis is split into four cases depending on the relative strength of the bids $A, B$ and $G$.}
\label{tab:payments}
\end{table}

\subsection{Application to Core-Selecting Payment Rules}

As explained in the preliminaries, many payment rules are defined by taking the reference point and projecting onto the point of the MRC that has the smallest distance in some metric such as $L_2$.
For all such \emph{minimum-revenue core-selecting (MRCS)} rules, we can characterize how manipulable they are (locally), only needing to know the sensitivity of the reference point and the position of the reference point relative to the individual rationality constraints.


\begin{lemma}
\label{lem:core_proj}

Let $p$ be a payment rule, and $\proj(p)$ a MRCS payment rule which uses $p$ as a reference point, i.e., it selects the point on the minimum-revenue core closest to $p$ according to the $L_c$ metric for some $c > 1$.

Then, we have that
\begin{equation}
\label{eq:lemma}
\frac{\partial \proj(p)_1}{\partial A}  = 
    \left\{ \begin{array}{ll} 
        1 & \mathrm{\ if\ } p_1 > p_2 - G + 2A \\
        0 & \mathrm{\ if\ } p_1 < p_2 + G - 2B \\
        \sens(p)_1 / 2 & \mathrm{\ otherwise\ } \\
    \end{array} \right.
\end{equation}

\end{lemma}

\begin{figure}
\centering
\includegraphics[width=.4\columnwidth]{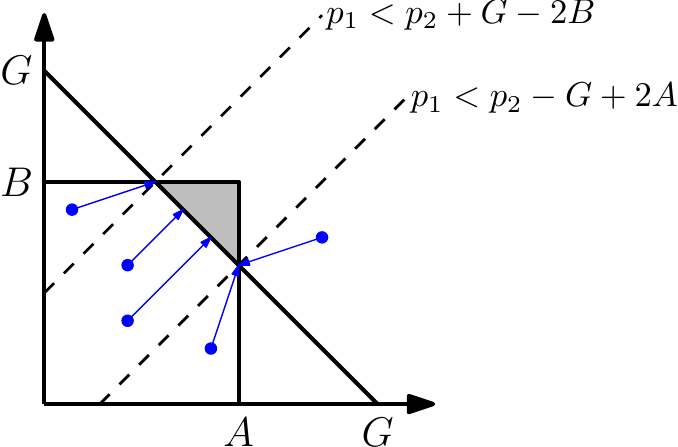}
\caption{The core in LLG. The projections of different reference points to the minimum-revenue core are shown in blue. The core projection reacts differently to changes in the reference point, depending on which side of the given inequalities the reference point is located.}
\label{fig:LLG_core}
\end{figure}

Condition \eqref{eq:lemma} of the lemma distinguishes three cases (see Figure~\ref{fig:LLG_core}).
In the first case, bidder 2's IR constraint is binding, so bidder 1 cannot affect his payment (locally).
In the second case, the projection to the MRC hits bidder 1's IR constraint, so any change to $b_1(A)$ immediately moves the payment by the same amount.
If neither of these two constraints are binding, the payment varies based on the sensitivity of the reference point.

\begin{proof}
In LLG, when the local bidders win, the minimum revenue core is given by the global constraint $p_1 + p_2 \geq G$, capped at each end by the local bidders' IR constraints, and there are no other core constraints to consider.

For any $L_c$ norm with $c > 1$, and any point $p$, the closest point to $p$ that lies on this global core constraint can be calculated as follows: we take the amount missing to exactly satisfy the constraint, i.e., $G - p_1 - p_2$, and split it evenly among the local bidders.
The inequalities given in the first two cases characterize exactly if this projection reaches the MRC, or if one of the two IR constraints become binding.

In the first two cases, the location of the reference point does not matter at all, as the intersection of the global constraint and the corresponding IR constraint is always chosen as the solution.
Thus, the derivative of the payment depends only on how a change in bidder 1's bid affects the IR constraint itself, which is 1 if it is the same bidders' IR constraint, and 0 if it is the other bidders'.

In the third case, bidder 1's payment, as argued above, is
\begin{equation}
\proj(p)_1 = p_1 + \frac{G - p_1 - p_2}{2} = \frac{G + p_1 - p_2}{2},
\end{equation}
and the derivative follows trivially.
\end{proof}

\begin{figure}
\centering

\begin{subfigure}{0.31\textwidth}
\centering
\includegraphics[width=.95\columnwidth]{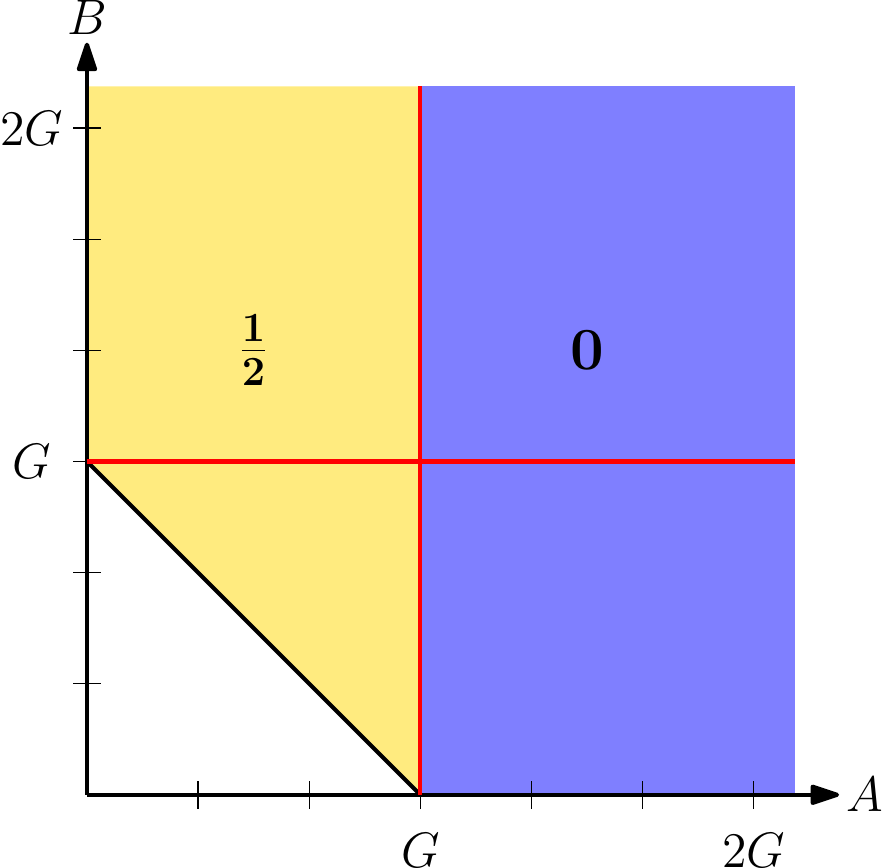}
\caption{Projection onto MRC from VCG (i.e., VCG-nearest).}
\label{fig:VCGN_deriv}
\end{subfigure}
\quad
\begin{subfigure}{0.31\textwidth}
\centering
\includegraphics[width=.95\columnwidth]{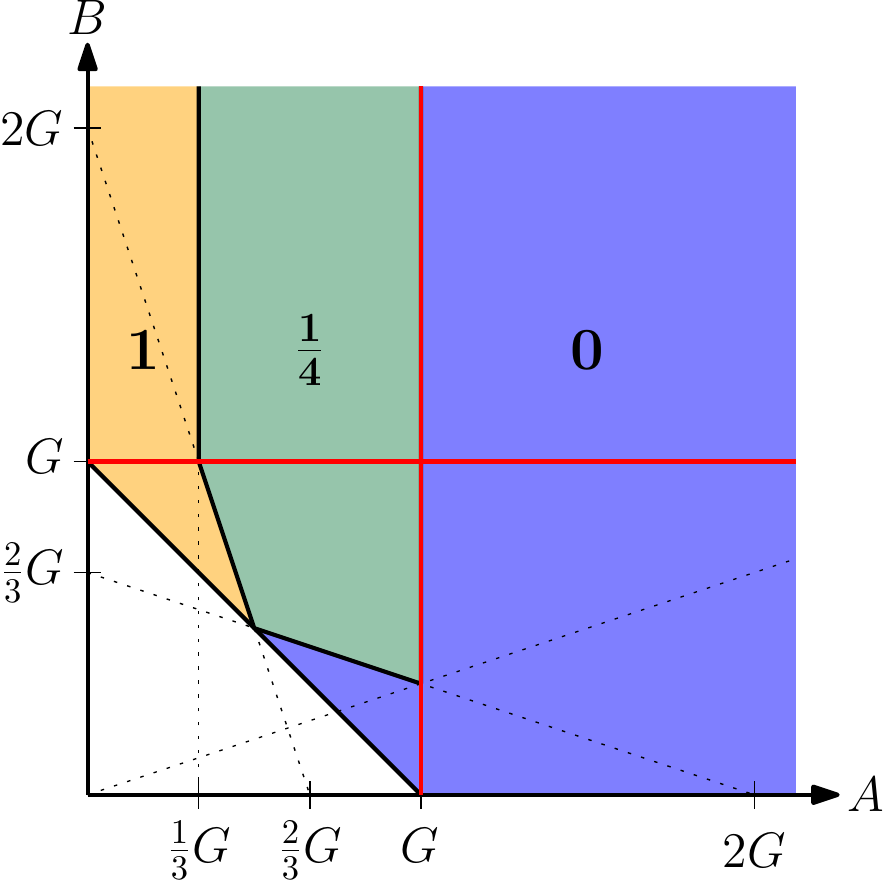}
\caption{Projection onto MRC from Shapley payments (without seller).}
\label{fig:Shapley_deriv}
\end{subfigure}
\quad
\begin{subfigure}{0.31\textwidth}
\centering
\includegraphics[width=.95\columnwidth]{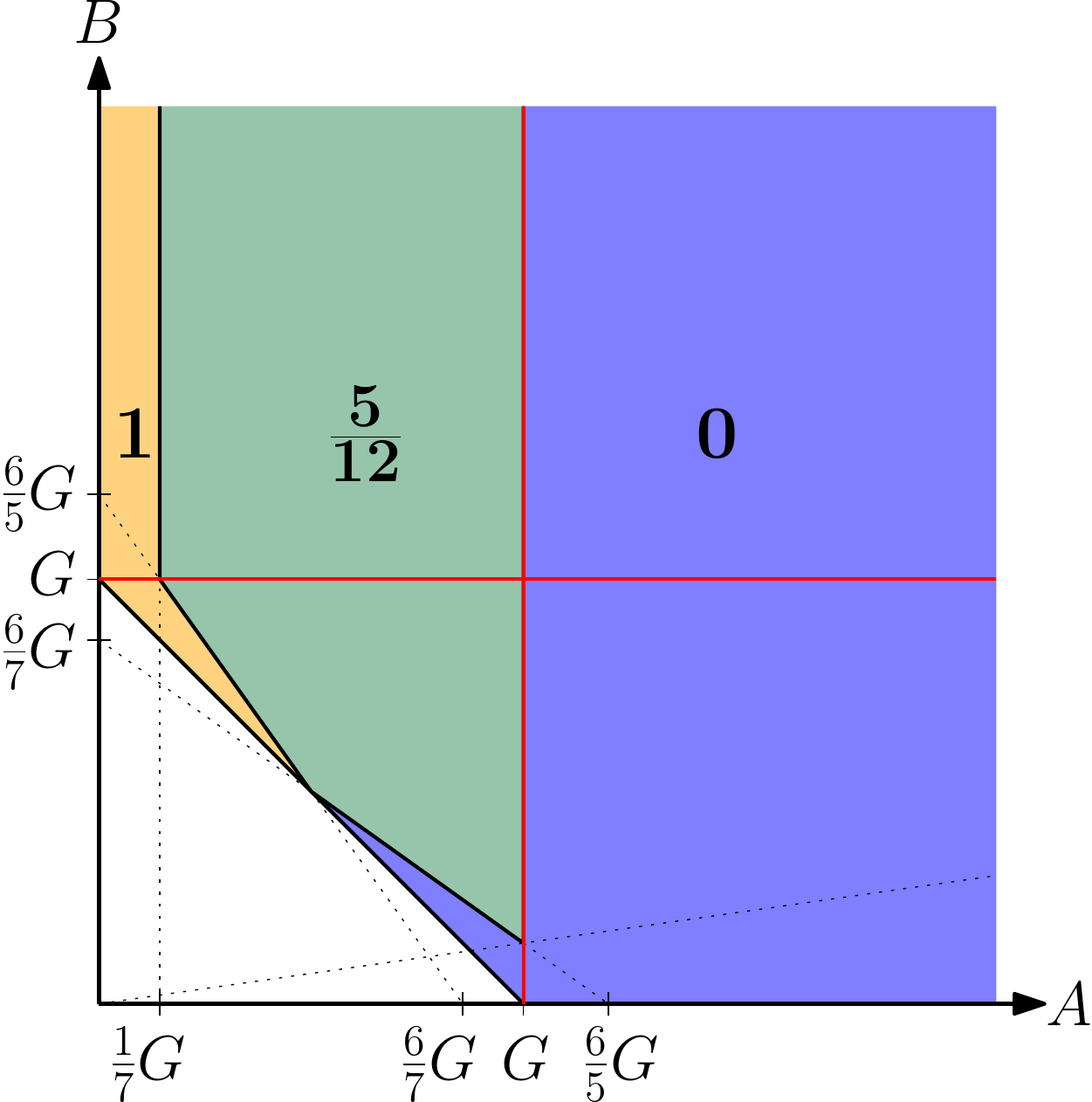}
\caption{Projection onto MRC from Shapley payments (with seller).}
\label{fig:Shapley2_deriv}
\end{subfigure}
\caption{Derivatives of three MRCS payment rules in the LLG domain. The MRC projections are done with any $L_c$ metric for $c > 1$. The derivatives are taken w.r.t. to bidder 1's bid $A$, and they depend on the bids $A$ and $B$ of both local bidders, as well as on the global bidder's bid $G$. The bidding space is divided into four cases as shown by the red lines. In the white area, the local bids sum up to less than $G$, so the global bidder wins both goods.}

\end{figure}

Based on Lemma~\ref{lem:core_proj}, we can now fully characterize many MRC payment rules in LLG.
In this note, we show three examples, VCG-nearest and Shapley-nearest (based on Shapley payments with and without the seller).

The VCG reference point always falls into case 3 of the Lemma, as can be easily verified by hand using the analytical expressions from Table~\ref{tab:payments}.
Thus, the derivative of VCG-nearest is exactly half the sensitivity of VCG, i.e., either $1/2$ or $0$, depending on bidder 1's bid (Figure~\ref{fig:VCGN_deriv}).

\

For Shapley-nearest, the inequalities from \eqref{eq:lemma} hold under the following conditions:

\vspace{3mm}
%
%
\begin{tabular}{|c|c|c||c|c|}
\hline
& \multicolumn{2}{c||}{Shapley payments (without seller)} & \multicolumn{2}{c|}{Shapley payments (with seller)} \\
\hline
 & $p_1 > p_2 - G + 2A$ & $p_1 < p_2 + G - 2B$  & $p_1 > p_2 - G + 2A$ & $p_1 < p_2 + G - 2B$ \\
\hline
weak locals & $3A + B < 2G$ & $A + 3B < 2G$  & $7A + 5B < 6G$ & $5A + 7B < 6G$ \\
strong local 1 & never holds & $A > 3B$ & never holds & $A > 7B$ \\
strong local 2 & $3A < G$ & never holds & $7A < G$ & never holds \\
strong locals &  never holds & never holds & never holds & never holds\\
\hline
\end{tabular}
\vspace{3mm}

The situation is summarized graphically in Figures~\ref{fig:Shapley_deriv}~and~\ref{fig:Shapley2_deriv}.
As we can see, in most cases, the local manipulability is much lower than for VCG-nearest, except when $A$ is very low compared to $G$.
This plausibly explains the experimental findings of \cite{lubin2018designing} that this payment rule has better incentive properties than VCG-nearest in LLG under incomplete information: while it is not monotonically better according to our metric, it is better when the corresponding bidder's value is high, and this case is more relevant to overall utility for that bidder.

\section{Discussion}

In this note, we presented some preliminary results pointing at a possible explanation why certain reference point rules are more manipulable than others.
Even though the VCG-nearest payment rule is widely used in practice \cite{ausubel2017practical}, our analysis shows that it has a relatively high susceptibility to being manipulated.
With the VCG point having a sensitivity of $1$, VCG-nearest behaves as an even mixture between the first price and VCG payment rules (in the appropriate sub-case), as has been shown previously \cite{bosshard2020cost}.

In contrast, we characterize payment rules based on the Shapley reference point as being surprisingly resistant to manipulations, at least in the LLG domain.

In future work, we aim to generalize the notion of sensitivity to larger domains of arbitrary dimensionality, to analyze the behavior of reference point rules in those domains, and to use those insights to better predict a bidder's behavior when acting under incomplete information.



\bibliographystyle{apalike}
\bibliography{/home/vbt/repositories/CCA/CApapers}

\begin{thebibliography}{}

\bibitem[Ausubel and Baranov, 2017]{ausubel2017practical}
Ausubel, L. and Baranov, O. (2017).
\newblock A practical guide to the combinatorial clock auction.
\newblock {\em The Economic Journal}, 127(605).

\bibitem[Ausubel and Milgrom, 2006]{ausubel2006lovely}
Ausubel, L. and Milgrom, P. (2006).
\newblock The lovely but lonely {V}ickrey auction.
\newblock In Cramton, P., Shoham, Y., and Steinberg, R., editors, {\em
  Combinatorial Auctions}. MIT Press.

\bibitem[Beck and Ott, 2013]{beck2013incentives}
Beck, M. and Ott, M. (2013).
\newblock Incentives for overbidding in minimum-revenue core-selecting
  auctions.
\newblock {\em Annual Conference 2013 (Duesseldorf): Competition Policy and
  Regulation in a Global Economic Order}.

\bibitem[Bosshard et~al., 2020]{bosshard2020JAIR}
Bosshard, V., B\"{u}nz, B., Lubin, B., and Seuken, S. (2020).
\newblock Computing bayes-nash equilibria in combinatorial auctions with
  verification.
\newblock {\em Journal of Artificial Intelligence Research}, 69:531--570.

\bibitem[Bosshard and Seuken, 2020]{bosshard2020cost}
Bosshard, V. and Seuken, S. (2020).
\newblock The cost of simple bidding in combinatorial auctions.
\newblock Working Paper, University of Zurich.

\bibitem[B\"{u}nz et~al., 2018]{lubin2018designing}
B\"{u}nz, B., Lubin, B., and Seuken, S. (2018).
\newblock Designing core-selecting payment rules: A computational search
  approach.
\newblock In {\em Proceedings of the 19th ACM Conference on Electronic Commerce
  (EC)}, Ithaca, NY, USA.

\bibitem[Meurer et~al., 2017]{meurer2017sympy}
Meurer, A., Smith, C.~P., Paprocki, M., {\v{C}}ert{\'\i}k, O., Kirpichev,
  S.~B., Rocklin, M., Kumar, A., Ivanov, S., Moore, J.~K., Singh, S., et~al.
  (2017).
\newblock Sympy: symbolic computing in python.
\newblock {\em PeerJ Computer Science}, 3:e103.

\bibitem[Shapley, 1953]{shapley1953value}
Shapley, L.~S. (1953).
\newblock A value for n-person games.
\newblock {\em Contributions to the Theory of Games}, 2(28):307--317.

\end{thebibliography}


\end{document}